\documentclass{article}
\usepackage{spconf,amsmath,graphicx}
\usepackage[shortlabels]{enumitem}
\usepackage[disable]{todonotes}
\usepackage{hyperref}

\title{Multi-Time-Scale Convolution for Emotion Recognition \\ from Speech Audio Signals       }
\name{Eric Guizzo, Tillman Weyde, Jack Barnett Leveson}
\address{Department of Computer Science \\
City, University of London \\
\texttt{\{eric.guizzo,t.e.weyde,jack.barnett-leveson\}@city.ac.uk}}

\begin{document}
%
\maketitle
\begin{abstract}
\todo{changed to be clearer}
Robustness against temporal variations is important for emotion recognition from speech audio, since emotion is expressed through complex spectral patterns that  can exhibit significant local dilation and compression on the time axis depending on speaker and context. 
To address this and potentially other tasks, we introduce the multi-time-scale (MTS) method to create flexibility towards temporal variations when analyzing  time-frequency representations of audio data. 
MTS extends convolutional neural networks with convolution kernels that are scaled and re-sampled along the time axis,
to increase temporal flexibility without increasing the number of trainable parameters compared to standard convolutional layers. 
We evaluate MTS and standard convolutional 
layers
in different architectures for 
emotion recognition from speech audio, 
using 4 
datasets of different sizes. 
The results show that the use of MTS layers consistently improves the generalization of networks of different capacity and depth, compared to standard convolution, 
especially on smaller datasets. 
\end{abstract}
\section{Introduction}
\label{sec:intro}

Convolutional Neural Networks (CNNs) have been extremely successful in recent years in a number of audio processing tasks, such as source separation, audio denoising, speech enhancement, speech and music transcription
\cite{
DBLP:conf/ismir/JanssonHMBKW17, 
enhancement1, 
speech_recognition1, 
music_transcription}. 
CNNs have also been extensively adopted for speech emotion recognition (SER) 
\todo{here reviewer 3 suggest to add 2 refs, but we have no space left for more references}
\cite{badshah2017speech, mao2014learning, trigeorgis2016adieu}. 

Convolutional networks benefit from translation invariance of the processing on the time and frequency axis of a spectrogram or other time-frequency representations. 
However, in speech there are also variations in the speed of  articulation between speakers and even of the same speaker in different situations.
Therefore, allowing for matching the same kernel in multiple versions that are scaled differently on the time axis is the main idea in this work. 
We implement this in a self-contained layer architecture, the multi-time-scale (MTS) convolution layer, which does not increase the number of parameters 
and increases the 
temporal flexibility in our networks compared to standard CNNs. 
Separate treatment of dimensions is useful for speech processing with time-frequency representations, as opposed to image processing, where scaling is normally applied to both dimensions. 
\todo{reviewer1 says:The work is presented solely as a machine learning contribution without specifying the context in which it might be suitable for the particular task of emotion recognition. We could add something here, but it would require a lot of space}

The contributions of our work are specifically:
\begin{itemize}
 \item 
 a convolution layer design for audio emotion recognition that learns locally-scale-invariant features in the time dimension
 \item 
 an evaluation of our approach to 4 emotion-labelled speech datasets with 4 different network architectures
 \item 
 an analysis of the experimental results, 
 confirming the effectiveness of the MTS approach. 
\end{itemize}

The remainder of the paper is organized as follows: Section~\ref{background} contains a brief review of relevant background literature, Section \ref{method} introduces the architecture of multi-time-scale convolution layer, Section~\ref{evaluation} presents the experimental results we obtained and Section~\ref{conclusions}  provides the conclusion of this paper.

\section{Related Work}
\label{background}

Scale-invariance in convolutional neural networks has been addressed in a number of 
ways. 
The most common approach for audio by far is data augmentation \cite{salamon2017deep,mcfee2015software}, 
which is frequently done by generating time-stretched variants of the training data. 
This procedure is usually part  of a pipeline of different transformations, as in \cite{augmentation1}, 
which has proven effective in various tasks. 
However, 
in this approach 
the different scales in the data need to be learned by different filters in the network. 
Therefore, greater network capacity is required and there is no guarantee that scale-invariance  is consistently achieved . 

Another strategy for 
scale-invariance in neural networks is to design it into the training and inference methods, so that it is applied consistently and without the need for additional training examples. 
There are many existing approaches to achieve this. 
The majority of them use a pyramidal structure, in which the scale is progressively narrowed along the network. 
\cite{felzenszwalb2009object} 
use parallel models trained with images at descending resolutions and then combine the obtained predictions as an ensemble model. 
\cite{szegedy2015going} achieve scale invariance with multiple loss functions, separately computed in layers with different resolutions within the network. 
Inception networks 
\cite{szegedy2015going} 
use parallel convolution layers with different filter sizes, matching features at different scales, 
but also increasing the number of variables in the network. 
\cite{wang2019elastic} propose a convolutional architecture, in which 
a 
scaling factor is learned by the network for every layer.

The majority of studies of 
scale-invariance in neural networks is focused on computer vision tasks. 
In the acoustic domain,  in addition to data augmentation techniques, 
scale-invariance can also be addressed through specific hard-coded transforms \cite{marchand2016scale} that are robust 
to some extent to scale variations. 
Nevertheless, since they are hard-coded, these methods 
need manual intervention %
and are usually highly task-specific, 
while embedding scale-invariance in the models provides a more generic solution that can be applied to multiple domains. 
The work of \cite{zhu2016learning} is an exception to this trend. 
They 
show that a network with $n$ identically-sized filters performs worse than a network with the same number of filters, but split in $3$  different sizes. 
%
Nevertheless, their models learn independent filters at different scales, 
increasing the number of free parameters. 

\textit{Locally scale-invariant convolutional neural networks}, as introduced for image recognition in 
\cite{SCALEINVARIANT}, 
are similar to our approach. 
This method consists of performing feature-extraction through multiple parallel convolution layers, whose outputs are locally merged through max-pooling. 
This produces a self-contained structure that can substitute a canonical convolution layer. 
The key feature of their approach is the possibility of matching a feature at multiple scales without increasing the number of free variables in the network. 
It permits introducing several re-scaled parallel branches at different points in the network, providing higher flexibility then pyramidal architectures.


\section{Method}
\label{method}
Our approach is similar to \cite{SCALEINVARIANT}, 
but specifically adapted to the audio domain, where we analyse 2D magnitude spectrograms of speech audio. 
Since the time and frequency dimensions are of different nature in this representation, we treat them  independently. 
Here, we focus on SER and address only time-scaling, while image processing techniques 
apply re-scaling to both dimensions with the same factor. 

The core of our architecture is the multi-time-scale convolution layer (MTS), a custom 2D-convolution layer that 
can replace 
a standard convolution layer in a CNN design. 
The main feature of MTS
is that it
uses multiple versions of the learned kernel that are re-sampled on the time axis and performs parallel convolutions with them. 
This method enables the network 
to detect patterns at multiple time scales.

\begin{figure}[tb]
  \centering
  \includegraphics[width=8.5cm]{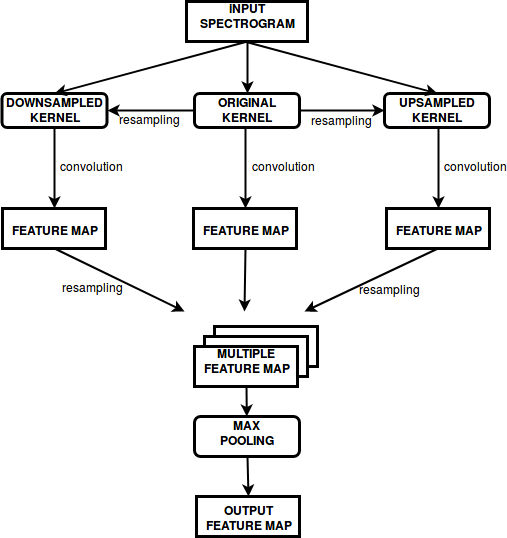}
\caption{Example architecture of a Multi-Time-Scale convolution layer with 3 scale factors.}
\label{fig:layer_architecture}
\end{figure}

Figure \ref{fig:layer_architecture} shows the architecture of one MTS layer with 3 parallel branches. 
In 
this example, 
the 2D spectrogram input, is convolved in parallel with the original kernel (in the center) and 2 time-stretched versions of the kernel (on both sides). 
The latter are generated by re-sampling the original kernel, applying linear interpolation. 
It is possible to independently apply different scaling factors for the 2 dimensions.
These parallel convolutions produce 3 different feature maps, matching the feature 
of 
the original kernel at 3 different time scales.
After this stage, the scaled feature maps are re-sampled again (applying linear interpolation) to match the shape of the original feature map. 
Then, a 3D max-pooling function is applied to merge the feature maps, selecting the scale with the maximal result in every time-frequency point. 
Therefore, the pooled feature map maintains the same dimension of the feature map generated by the original kernel.
During the training we average the weights of the original kernel and its scaled versions after each update.
There is no constraint by design on the number of parallel branches that can be added to a MTS layer 
and MTS layers with different numbers of branches can be placed at various positions in the network.
It is possible to fine-tune the scaling factors layer-by-layer.
This approach provides a high degree of flexibility in the network design and enables scale invariance without increasing the number of free parameters. 
We have implemented this method in PyTorch as open source\footnote{\url{https://github.com/ericguizzo/multi_time_scale}}.

\todo{this explains better the kernel rescaling, as suggested by reviewer2}
Our method is different from \cite{SCALEINVARIANT} in that it re-scales only one dimension and that we re-sample the kernels.
Although re-sampling the data or kernel is equivalent up to numerical variations, our method is somewhat more efficient.
Moreover, \cite{SCALEINVARIANT} augment test data by re-scaling.
At least for SER tasks, we believe that this practice would not give a good estimate of the generalization capabilities of the models and thus we test without  augmentation.


\section{Evaluation}
\label{evaluation}
\todo{changed different to benchmark. Can we say benchmark even though TESS is not commonly used?}
We have evaluated the performance of MTS on 4 benchmark datasets for speech emotion recognition:

\begin{enumerate}
  \item EMODB, a database of German emotional speech  \cite{burkhardt2005database}. 10 speakers, German language, 535 utterances, 25 min of audio, 7 emotion labels: angry, bored, disgusted, anxious/fearful, happy, sad. 
  Actors pronounce 10 different sentences which could be used in everyday communication.
  
  \item RAVDESS, the Ryerson Audio Visual Database of Emotional Speech and Song \cite{ravdess}. 24 speakers, English language, 2542 utterances, 2:47 hours of audio, 8 emotion labels: happy, sad, angry, fearful, surprised, disgusted, calm, neutral. 
  Actors pronounce 2 sentences: ``Kids are talking by the door'' and ``Dogs are sitting by the door''. 
  
  \item TESS, the Toronto Emotional Speech Set \cite{dupuis2010toronto}. 2 speakers, English language, 2800 utterances, 1:36 hours of audio, 7 emotion labels: happy, sad, angry, disgusted, neutral, pleasant surprise, fearful. 
  Actors say ``Say the word ...'' followed by 200 different words.
   
  \item IEMOCAP, the Interactive Emotional Dyadic Motion Capture Database  \cite{busso2008iemocap}. 5 speakers, English language, 7529 utterances, 9:32 hours of audio, 10 emotion labels: neutral, angry, happy, excited, sad, frustrated, fearful, surprised, disgusted, other. 
  Actors perform improvisations or scripted scenarios on defined topics.
\end{enumerate}

For each dataset we keep only the audio information and the emotion labels, discarding any other types of data. 
We also discard the ``song'' data from RAVDESS.
\todo{changed this sentence to make clear that IEMOCAP was the only inbalanced dataset}
IEMOCAP is the only highly inbalanced dataset, therefore we removed the rarest labels from it, keeping only neutral, angry, happy and sad samples. 
Every sound file is pre-processed in 3 consecutive stages: re-sampling to 16 kHz, 
Short-Time Fourier Transform
and normalization. 
For EMODB, RAVDESS and TESS datasets every file is zero-padded to obtain equally-sized data. 
Since the IEMOCAP dataset contains longer recordings we segmented them into 4-second frames with 2-second overlap.
The STFT is computed using 20 ms sliding windows with 10 ms overlap. 
Then, we normalize the magnitude spectra to zero mean and unit standard deviation. 

\begin{table*}[tbh]
  \caption{Accuracy results for all datasets. N ist the number of audio recordings per dataset. A1-4 are the network architectures. The usage factors relate to scaling factors in the same row. The best results per dataset are highlighted in bold font.}
  \centering
  \vspace{.5mm}
  \begin{tabular}{|l|l|l|l|l|l|l|l|l|}
    \hline
    Dataset & N &Type &A1   &A2  &A3  &A4  & Best scale factors  & Use of parallel branches\\
    \hline
    EMODB  &535 &Standard   &64.3   &66.26   &66.91       &62.75  & n/a  & n/a\\
    &535 &MTS   &66.5   &\textbf{70.97}   &70.68 
       &66.28  & 0.7, 1, 1.428  &0.47, 0.05, 0.48\\
    \hline
    RAVDESS  &1440 &Standard  &42.09   &39,84    &42.56     &47.41   & n/a & n/a \\
    &1440 &MTS &47.85   &44.95    &51.32 
        &\textbf{55.85}  &0.5, 1, 2    &0.45, 0.06, 0.49\\
    \hline
    TESS  &2800 & Standard  &47.45  &49.6   &50.61  &40.78  & n/a  & n/a\\
    &2800 & MTS  &51.76   &48.75   &\textbf{53.05}  &51.71  &0.5, 0.7, 1, 1.428, 2.  &0.41, 0.04, 0.05, 0.07, 0.43\\
    \hline
    IEMOCAP  &5531 & Standard  &48.93   &50.48    &49.0   &54.96  & n/a & n/a \\
     &5531 &MTS  &49.0   &50.84    &49.86 
       &\textbf{55.01} &0.5, 0.7, 1, 1.428, 2 &0.39, 0.04, 0.04, 0.05, 0.48\\
    \hline
  \end{tabular}
  
  \label{table:results}
\end{table*}

We divide every dataset using approximately 70\% of the data as training, 20\% for validation and 10\% as test set. 
Furthermore, we perform every experiment with 4-fold cross-validation.
We 
make sure that 
samples 
from the same speaker appear only in the same set, in order to get a meaningful measure of the models' capability to generalize to new speakers, because new speakers are likely to produce patterns at different speeds. 
For this and other reasons, our results are not directly comparable to most published results.
Many results are computed with randomly-split training, validation and test sets, without separating speakers, as in
\cite{zeng2019spectrogram}. 
Many rely on different preprocessing 
\cite{shegokar2016continuous, schuller2005speaker}, 
on different architectures 
\cite{shegokar2016continuous}
or use multi-modal features rather than only audio 
\cite{schuller2005speaker}. 
Rather than aiming at a state-of-art classification accuracy for these datasets, we focus on evaluating the performance of MTS layers compared to standard convolution with the same number of channels, 
i.e. without increasing the number of trainable variables.
Therefore, we arranged our experiments in order to obtain consistent results within our set-up, with the same conditions for all datasets.
We perform this comparison for 4 different CNN architectures with different capacity: 
\begin{enumerate}[start=1,label={\bfseries A\arabic*:}]
  \item Convolution (1 channel, [10,5] kernel) - fully connected (200 neurons) - fully connected output layer.
  \item Convolution (10 channels, [10,5] kernel) - fully connected (200 neurons) - fully connected output layer.
  \item Convolution (10 channels, [10,5] kernel) - max pooling ([2,2] kernel) - convolution (10 channels, [10,5] kernel - fully connected (200 neurons) - fully connected output layer.
  \item AlexNet: 5 convolutions and max pooling,  2 fully connected layers. See \cite{krizhevsky2012imagenet} for a detailed description. 
\end{enumerate}
The kernel dimensions above are in the form [time,frequency].
The activation function is ReLU for hidden and softmax for output units. 
In all experiments we use the ADAM optimizer with L2 regularization and Cross Entropy loss. 
We perform a grid search to find the best regularization parameter. 
We train for a maximum of 500 epochs, applying early stopping with 10 epochs patience 
for validation loss improvement.
In architectures A1, A2 and A3, MTS is applied to all convolutional layers, while in A4 only the first 2 layers are augmented with MTS. 
We tested MTS with 3, 5 and 7 parallel branches, using logarithmically spaced scale factors  
in these combinations: (0.25, 1, 4), (0.5, 1, 2), (0.7, 1, 1.428), (0.8, 1, 1.25), (0.9, 1, 1.111), (0.95, 1, 1.053), (0.25, 0.5, 1, 2, 4), (0.5, 0.7, 1, 1.428, 2), (0.8, 0.9, 1, 1.111, 1.25), (0.25, 0.5, 0.7, 1, 1.428, 2, 4), (0.7, 0.8, 0.9, 1, 1.111, 1.25, 1.428). 
\todo{added this line to make the use of stretch factors clearer, suggested by reviewer 2}
In each experiment, we apply the same combination of stretch factors to all MTS-enabled layers.

Table \ref{table:results} shows the results we obtained for all datasets and all architectures.  
The first 3 columns show 
the dataset, total number of data points 
and the type of convolution layer(s). 
Columns A1-A4 show the mean test accuracy across folds obtained with each architecture (as listed above). 
The last 2 columns refer to the MTS model with the best accuracy in a row and show the scaling factors applied to each parallel branch of MTS and their average percentage of use.

The results clearly show that MTS consistently improves the generalization for all datasets. 
We reach a maximum improvement of 8.04 percentage points (RAVDESS) and with an average of 3.78 with a standard deviation of 3.45 across all datasets and architectures.
For all 
model/architecture combinations except one (A2 with TESS), MTS outperforms standard convolution.
\todo{shortened, see comment}
We performed a two-sided Wilcoxon signed-rank test 
comparing the standard and MTS results, which shows 
statistical significance with $p < 0.001$. 
The mean improvement is  higher for the smaller  datasets, 
which confirms that 
enabling pattern recognition at different time scales with MTS improves generalisation. 
Considering the general scarcity of emotion-labelled speech data, this is a desirable feature for SER applications.

The best performing models on different datasets used different combinations of scaling factors. 
In particular, for the smaller datasets applying only 3 factors gives the best results. 
Architectures with 5 parallel branches perform better for the larger datasets. 
MTS models tend to use mostly 2 scale factors (see last column of table \ref{table:results}). 
In every case, at least 2 parallel branches give a high contribution, confirming that MTS is actually matching patterns at multiple time-scales.

We found that MTS is more effective at larger kernel sizes. 
In an experiment with an MTS version of ResNet18,
where most kernels are very small (3x3), we achieved no improvement with MTS. 

Training a MTS-enabled network  generally takes longer than a standard CNN.
In a test with architecture A2, it took on average 1.3 times longer per epoch to train MTS models with 3 branches and 1.52 times longer for MTS models 5 branches.
Moreover, MTS networks need on average more epochs to converge (27.85 vs 32.26 epochs for CNN vs MTS average overall). 
\newpage
\todo{better to cut something after this point to avoid describing the  list in the end of the previous column and starting the actual list in the new column}
We also tested modified variants of MTS: 
\begin{itemize}
 \item Applying a penalty to the re-sampled feature maps, to give the model a preference for the unscaled kernel.
 \item Performing the training using standard convolution layers and substitute them with MTS layers with shared weights only at inference time.
 \item 
 Concatenating the used scaling factor for each time-frequency point to the output feature map of an MTS layer. 

\end{itemize}
Each of these modifications  reduced the performance of MTS models. 
Therefore, we kept the  simplest variant described above.

\section{Conclusions}
\label{conclusions}

In this paper, we propose a multi-time-scale convolution layer (MTS) for CNNs applied to audio analysis, specifically emotion recognition from speech. 
The MTS performs parallel 2D-convolutions using a  standard kernel and its re-sampled versions to match  patterns at different time scales. 
This method enables the network to learn to some extent time-invariant features without increasing its number of trainable parameters or the number of training examples. 
We evaluated our approach on speech emotion recognition with unknown speakers, using 4 different datasets and applying it to networks of different size and structure. 
We found a consistent and statistically significant improvement in test accuracy across all datasets and models, up to 8.04 percentage points for RAVDESS and on average 3.78 
across all datasets and architectures.
MTS is particularly effective on smaller datasets, which  makes MTS well suited for Speech Emotion Recognition where labelled data is scarce.

As future developments we intend to test more extensively the effectiveness of MTS with larger datasets and explore more architectures and different resampling techniques. 
Furthermore, we are going to apply the concept of MTS in the
context of convolution-based generative models, 
\todo{ cut 'as Variational Autoencoders and Generative Adersarial Networks' to save one more line}
extending our multi-branch approach also to transposed convolutions.

\vspace{0.5cm}

\bibliographystyle{IEEEbib}
\bibliography{ICASSP2020}

\begin{thebibliography}{10}

\bibitem{DBLP:conf/ismir/JanssonHMBKW17}
Andreas Jansson, Eric~J. Humphrey, Nicola Montecchio, Rachel~M. Bittner, Aparna
  Kumar, and Tillman Weyde,
\newblock ``Singing voice separation with deep u-net convolutional networks,''
\newblock in {\em ISMIR}, 2017, pp. 745--751.

\bibitem{enhancement1}
Szu-Wei Fu, Yu~Tsao, and Xugang Lu,
\newblock ``Snr-aware convolutional neural network modeling for speech
  enhancement,''
\newblock in {\em Interspeech}, 2016, pp. 3768--3772.

\bibitem{speech_recognition1}
Dimitri Palaz, Ronan Collobert, et~al.,
\newblock ``Analysis of cnn-based speech recognition system using raw speech as
  input,''
\newblock Tech. {R}ep., Idiap, 2015.

\bibitem{music_transcription}
Rachel~M Bittner, Brian McFee, Justin Salamon, Peter Li, and Juan~Pablo Bello,
\newblock ``Deep salience representations for f0 estimation in polyphonic
  music,''
\newblock in {\em ISMIR}, 2017, pp. 63--70.

\bibitem{badshah2017speech}
Abdul~Malik Badshah, Jamil Ahmad, Nasir Rahim, and Sung~Wook Baik,
\newblock ``Speech emotion recognition from spectrograms with deep
  convolutional neural network,''
\newblock in {\em PlatCon}, 2017, pp. 1--5.

\bibitem{mao2014learning}
Qirong Mao, Ming Dong, Zhengwei Huang, and Yongzhao Zhan,
\newblock ``Learning salient features for speech emotion recognition using
  convolutional neural networks,''
\newblock {\em IEEE Multimedia}, vol. 16, no. 8, pp. 2203--2213, 2014.

\bibitem{trigeorgis2016adieu}
George Trigeorgis, Fabien Ringeval, Raymond Brueckner, Erik Marchi, Mihalis~A
  Nicolaou, Bj{\"o}rn Schuller, and Stefanos Zafeiriou,
\newblock ``Adieu features? end-to-end speech emotion recognition using a deep
  convolutional recurrent network,''
\newblock in {\em ICASSP}, 2016, pp. 5200--5204.

\bibitem{salamon2017deep}
Justin Salamon and Juan~Pablo Bello,
\newblock ``Deep convolutional neural networks and data augmentation for
  environmental sound classification,''
\newblock {\em IEEE Signal Processing Letters}, vol. 24, no. 3, pp. 279--283,
  2017.

\bibitem{mcfee2015software}
Brian McFee, Eric~J Humphrey, and Juan~Pablo Bello,
\newblock ``A software framework for musical data augmentation,''
\newblock in {\em ISMIR}, 2015, pp. 248--254.

\bibitem{augmentation1}
Jan Schl{\"u}ter and Thomas Grill,
\newblock ``Exploring data augmentation for improved singing voice detection
  with neural networks,''
\newblock in {\em ISMIR}, 2015, pp. 121--126.

\bibitem{felzenszwalb2009object}
Pedro~F Felzenszwalb, Ross~B Girshick, David McAllester, and Deva Ramanan,
\newblock ``Object detection with discriminatively trained part-based models,''
\newblock {\em IEEE PAMI}, vol. 32, no. 9, pp. 1627--1645, 2009.

\bibitem{szegedy2015going}
Christian Szegedy, Wei Liu, Yangqing Jia, Pierre Sermanet, Scott Reed, Dragomir
  Anguelov, Dumitru Erhan, Vincent Vanhoucke, and Andrew Rabinovich,
\newblock ``Going deeper with convolutions,''
\newblock in {\em CVPR}, 2015, pp. 1--9.

\bibitem{wang2019elastic}
Huiyu Wang, Aniruddha Kembhavi, Ali Farhadi, Alan~L Yuille, and Mohammad
  Rastegari,
\newblock ``Elastic: Improving cnns with dynamic scaling policies,''
\newblock in {\em CVPR}, 2019, pp. 2258--2267.

\bibitem{marchand2016scale}
Ugo Marchand and Geoffroy Peeters,
\newblock ``Scale and shift invariant time/frequency representation using
  auditory statistics: Application to rhythm description,''
\newblock in {\em IEEE MLSP}, 2016, pp. 1--6.

\bibitem{zhu2016learning}
Zhenyao Zhu, Jesse~H Engel, and Awni Hannun,
\newblock ``Learning multiscale features directly from waveforms,''
\newblock {\em arXiv:1603.09509}, 2016.

\bibitem{SCALEINVARIANT}
Angjoo Kanazawa, Abhishek Sharma, and David Jacobs,
\newblock ``Locally scale-invariant convolutional neural networks,''
\newblock in {\em NIPS Deep Learning and Representation Learning Workshop}, 12
  2014.

\bibitem{burkhardt2005database}
Felix Burkhardt, Astrid Paeschke, Miriam Rolfes, Walter~F Sendlmeier, and
  Benjamin Weiss,
\newblock ``A database of german emotional speech,''
\newblock in {\em Eurospeech}, 2005.

\bibitem{ravdess}
Steven~R Livingstone and Frank~A Russo,
\newblock ``The ryerson audio-visual database of emotional speech and song
  (ravdess): A dynamic, multimodal set of facial and vocal expressions in north
  american english,''
\newblock {\em PloS one}, vol. 13, no. 5, pp. e0196391, 2018.

\bibitem{dupuis2010toronto}
Kate Dupuis and M~Kathleen Pichora-Fuller,
\newblock {\em Toronto Emotional Speech Set (TESS)},
\newblock University of Toronto, Psychology Department, 2010.

\bibitem{busso2008iemocap}
Carlos Busso, Murtaza Bulut, Chi-Chun Lee, Abe Kazemzadeh, Emily Mower, Samuel
  Kim, Jeannette~N Chang, Sungbok Lee, and Shrikanth~S Narayanan,
\newblock ``Iemocap: Interactive emotional dyadic motion capture database,''
\newblock {\em Language resources and evaluation}, vol. 42, no. 4, pp. 335,
  2008.

\bibitem{zeng2019spectrogram}
Yuni Zeng, Hua Mao, Dezhong Peng, and Zhang Yi,
\newblock ``Spectrogram based multi-task audio classification,''
\newblock {\em Multimedia Tools and Applications}, vol. 78, no. 3, pp.
  3705--3722, 2019.

\bibitem{shegokar2016continuous}
Pankaj Shegokar and Pradip Sircar,
\newblock ``Continuous wavelet transform based speech emotion recognition,''
\newblock in {\em IEEE ICSPCS}, 2016, pp. 1--8.

\bibitem{schuller2005speaker}
Bj{\"o}rn Schuller, Ronald M{\"u}ller, Manfred Lang, and Gerhard Rigoll,
\newblock ``Speaker independent emotion recognition by early fusion of acoustic
  and linguistic features within ensembles,''
\newblock in {\em Eurospeech}, 2005.

\bibitem{krizhevsky2012imagenet}
Alex Krizhevsky, Ilya Sutskever, and Geoffrey~E Hinton,
\newblock ``Imagenet classification with deep convolutional neural networks,''
\newblock in {\em NIPS}, 2012, pp. 1097--1105.

\end{thebibliography}

\end{document}